\documentclass[manuscript,screen]{acmart}
\AtBeginDocument{%
  }

\setcopyright{acmlicensed}
\copyrightyear{2025}
\acmYear{2025}
\acmDOI{XXXXXXX.XXXXXXX}
\acmConference[Conference acronym 'XX]{Make sure to enter the correct
  conference title from your rights confirmation email}{June 03--05,
  2025}{Woodstock, NY}
\acmISBN{978-1-4503-XXXX-X/2025/06}

\graphicspath{{./figs/}}
\DeclareGraphicsExtensions{.pdf,.jpeg,.png,.eps,.svg}

\usepackage{subcaption}
\usepackage{booktabs, tabularx}
\usepackage{multirow}
\usepackage{xcolor}
\usepackage{xspace}

\usepackage{algorithm}
\usepackage{algpseudocode}
\floatname{algorithm}{Procedure}

\usepackage{enumerate}
\usepackage{fancybox}  
\usepackage{multirow}
\usepackage{pifont}
\newcommand{\cmark}{\ding{51}}
\newcommand{\xmark}{\ding{55}}

\newcommand{\enquote}[1]{``#1''}

\newcommand{\dataset}{\textsc{AIDev}} 
\newcommand{\datasetpop}{\textsc{AIDev-pop}}
\newcommand{\agenticai}{{Autonomous Coding Agent}\xspace}
\newcommand{\agenticais}{{Autonomous Coding Agents}\xspace}
\newcommand{\codex}{\textsc{OpenAI Codex}\xspace}
\newcommand{\copilot}{\textsc{GitHub Copilot}\xspace}
\newcommand{\devin}{\textsc{Devin}\xspace}
\newcommand{\cursor}{\textsc{Cursor}\xspace}
\newcommand{\claude}{\textsc{Claude Code}\xspace}
\newcommand{\tablefmt}[1]{\texttt{#1}}
\newcommand{\botfmt}[1]{\texttt{#1}}

\newcommand{\agentprs}{Agentic-PRs\xspace}

\newcommand{\humanprs}{Human-PRs\xspace}

\newcounter{findingctr}
\newcounter{researchctr}

\newcommand{\finding}[1]{%
  \refstepcounter{findingctr}%
  \textbf{Finding~\#\thefindingctr: #1}%
}

\newcommand{\researchdir}[1]{%
  \refstepcounter{researchctr}%
  \textbf{Research Direction~\#\theresearchctr: #1}%
}

\DeclareTextFontCommand{\emp}{\bfseries}

\usepackage[most]{tcolorbox}
\definecolor{custom-gray}{cmyk}{0, 0, 0, 0.7, 1.00}

\newtcolorbox{Summary}[2][]{
top=0.15in,
fonttitle=\bfseries,
colbacktitle=custom-gray,
colback=gray!5,
colframe=gray!40!black,
enhanced,
attach boxed title to top left={xshift=1.5em,yshift=-\tcboxedtitleheight/2},
boxed title style={size=small,colback=custom-gray},
drop shadow={black!50!white},
title=#2,#1}

\begin{document}

\title[The Rise of AI Teammates in Software Engineering (SE) 3.0]{The Rise of AI Teammates in Software Engineering (SE) 3.0: How Autonomous Coding Agents Are Reshaping Software Engineering}

\author{Hao Li}
\email{hao.li@queensu.ca}
\orcid{0000-0003-4468-5972}
\affiliation{%
  \institution{Queen's University}
  \city{Kingston}
  \state{ON}
  \country{Canada}
}

\author{Haoxiang Zhang}
\email{haoxiang.zhang@queensu.ca}
\orcid{0000-0002-3921-1724}
\affiliation{%
  \institution{Queen's University}
  \city{Kingston}
  \state{ON}
  \country{Canada}
}

\author{Ahmed E. Hassan}
\email{ahmed@cs.queensu.ca}
\orcid{0000-0001-7749-5513}
\affiliation{%
  \institution{Queen's University}
  \city{Kingston}
  \state{ON}
  \country{Canada}
}



\begin{abstract}

The future of software engineering---SE 3.0---is already unfolding with the rise of AI teammates: autonomous, goal-driven systems that collaborate with human developers in real-world workflows. Autonomous coding agents are among the most transformative of these systems. Such agents are now actively initiating, reviewing, and evolving code at scale throughout the open source ecosystem. In this paper, we introduce AIDev, the first large-scale dataset capturing how such autonomous coding agents operate in the wild. Spanning over 456,000 pull requests authored by five leading autonomous agents---OpenAI Codex, Devin, GitHub Copilot, Cursor, and Claude Code---across 61,000 repositories and involving 47,000 developers, AIDev provides an unprecedented empirical foundation for studying the integration of autonomous teammates into modern software development.

Unlike prior work that has largely theorized the rise of AI-native software engineering, AIDev offers concrete, structured, and open data that can power future research in benchmarking, agent readiness, optimization avenues, collaboration modeling, and AI governance. The dataset includes rich metadata on pull requests (PR), authorship, review timelines, code changes, and integration outcomes---enabling researchers to explore questions that existing synthetic benchmarks like SWE-bench cannot answer. For example, although agents frequently outperform humans in speed, our analysis shows their pull requests are accepted less frequently, revealing a stark gap between benchmark performance and real-world trust and utility. Moreover, while agents can massively accelerate code submission---one developer submitted as many Agentic-PRs in three days as they submitted without Agentic help in the previous three years---these contributions tend to be structurally simpler (quantified through traditional code complexity metrics).

We envision AIDev as a living resource: extensible, analyzable, and immediately usable by the software engineering and AI communities. By grounding the vision and planning of SE 3.0 in real-world evidence, we hope to unlock a new generation of research into AI-native SE workflows, and lay the empirical groundwork for building and governing the next wave of symbiotic human–AI collaboration in SE. The AIDev dataset is publicly available at \url{https://github.com/SAILResearch/AI_Teammates_in_SE3}.

\end{abstract}

\ccsdesc[500]{Software and its engineering~Software development techniques}
\ccsdesc[500]{Software and its engineering~Collaboration in software development}
\ccsdesc[500]{Computing methodologies~Artificial intelligence}
\ccsdesc[500]{Software and its engineering~Software creation and management}

\keywords{AI Agent, Agentic AI, Coding Agent, Agentic Coding, Software Engineering Agent}

\maketitle

\section{Introduction}\label{sec:introduction}

\begin{flushright}
  \itshape
  ``Software is eating the world, but AI is going to eat software.''\\
  --- Jensen Huang (CEO of NVIDIA)
\end{flushright}

\begin{figure}[h]
  \centering
  
  \begin{subfigure}[t]{0.48\linewidth}
    \centering
    \includegraphics[width=\linewidth]{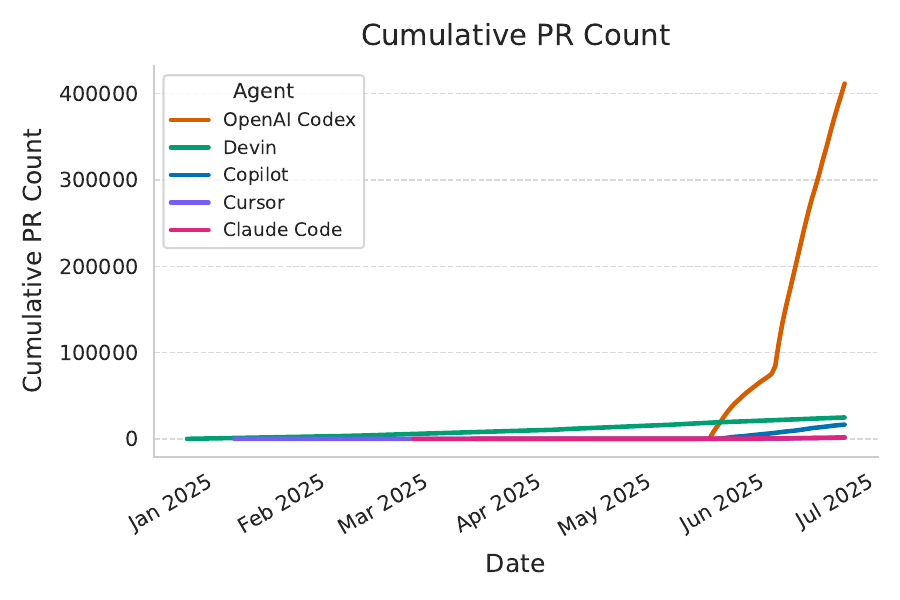}
    \caption{\dataset{} (full dataset)}
    \label{fig:pr_cumulative}
  \end{subfigure}
  \hfill
  \begin{subfigure}[t]{0.48\linewidth}
    \centering
    \includegraphics[width=\linewidth]{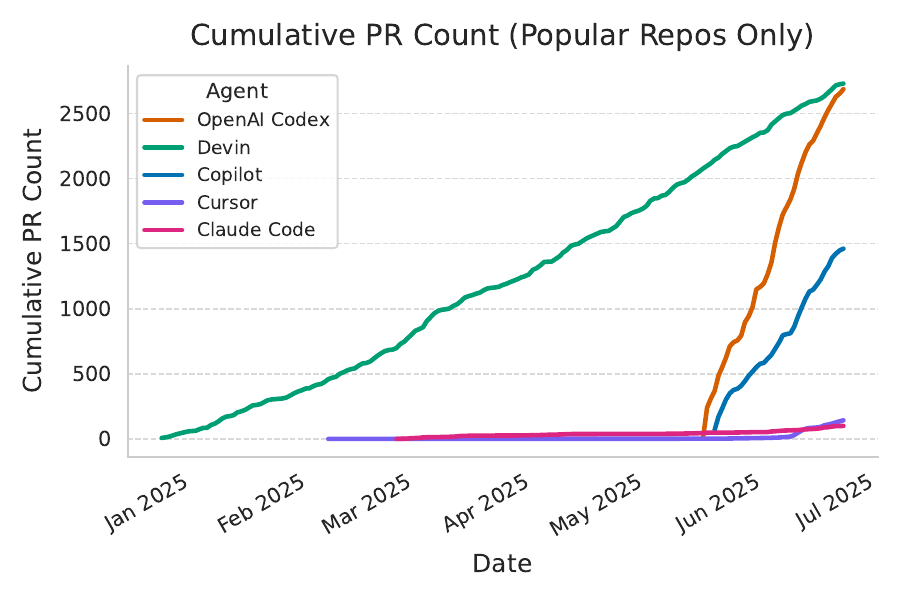}
    \caption{\datasetpop{} (repositories with ${>}500$ stars)}
    \label{fig:pr_cumulative_popular}
  \end{subfigure}
  
  \caption{Cumulative PR volume by \agenticais, shown across all repositories (\dataset{}) and popular repositories (\datasetpop{}).}
  \label{fig:pr_cumulative_combined}
\end{figure}

Software Engineering~(SE) is entering a new era, one increasingly shaped by AI Teammates: autonomous, task-driven agents capable of completing complex SE tasks such as feature development, debugging, and code review with minimal human oversight. Among the most transformative of these agents are \agenticais. While early AI coding assistants operated primarily in an autocomplete or co-pilot role, today's \agenticais are beginning to function as AI teammates who initiate pull requests and engage in feedback loops; in turn reshaping software development workflows. What was once a futuristic vision~\cite{hassan_se3_2024} is now unfolding rapidly: these AI teammates are actively contributing to codebases at scale. \agenticais are becoming routine actors in software development workflows, participating in thousands of pull requests (PRs)\footnote{From now on, we refer to such pull requests as \agentprs.} daily. Based on our collected dataset, \codex~\cite{openai_codex} alone has created over 400,000 PRs in open-source GitHub repositories in less than two months since its release in May 2025 (see Figure~\ref{fig:pr_cumulative}).

Much of the existing literature has anticipated this shift. Researchers have speculated about the rise of AI-native SE (SE 3.0)~\cite{hassan_se3_2024}, examined challenges around trust and reliability in human-AI interactions~\cite{hassan_aiware_2024}, and proposed system architectures that blend human creativity with AI-native automation. However, these discussions were primarily conceptual or forward-looking. So far, we lacked empirical grounding for how such AI teammates (e.g., \agenticais) would actually operate in the wild? How would \agenticais affect developer productivity and review latency? Do \agenticais produce \textit{mergeable} code? And how are human teams adapting to the presence of \agenticais?

\textbf{Our paper is the first work to demonstrate that the era of \agenticai in SE is not an impending future, but an unfolding reality.}
To support this claim, we introduce \dataset{}, a new large-scale dataset comprised of 456,535 Agentic PRs created by five leading \agenticais: \codex, \copilot, \devin, \cursor, and \claude on GitHub. Spanning 47,303 developers and 61,453 repositories, \dataset{} provides the first systematic look at how these AI teammates are being used, how they interact with human collaborators, and the nature of their contributions. Unlike existing codeLLM benchmarks (e.g., SWE-bench), which tend to rely on static and curated datasets, and fail to capture the messy and emergent nature of real-world SE; \dataset{} offers a dynamic, large-scale perspective grounded in real-life human-AI collaborations. We designed \dataset{} to serve both as an observational lens and a catalyst for future research, reflecting how \agenticais are used in practice.

\textbf{We envision the \dataset{} dataset as a foundation for the next frontier of software engineering research.} Researchers can use it to study human-AI cooperation, longitudinal productivity trends, \agenticai review behaviours, and emerging governance challenges around AI autonomy in real-world software projects. By grounding the vision of SE 3.0~\cite{hassan_se3_2024} in large-scale empirical evidence, \dataset{} bridges the gap between an unpredictable, fast-evolving future and ongoing real-world software engineering practices.

The main contributions of our paper are as follows:

\begin{itemize}
    \item We curate and release \dataset{},\footnote{\url{https://github.com/SAILResearch/AI_Teammates_in_SE3}} a large-scale dataset of 456,535 Agentic-PRs created by \agenticais across 61,453 GitHub repositories, capturing the breadth and diversity of \agenticais activity in the open-source software ecosystem.
    \item We present three case studies to demonstrate the value of \dataset{} and highlight the following key findings:
    \begin{itemize}
        \item Although \agenticais like \copilot can complete half of their PRs within 13 minutes, we find that their PRs are less likely to be accepted than those of human developers, indicating quality gaps in \agentprs and deviation from traditional benchmark results.
        \item PRs authored by \codex are reviewed and merged more quickly than human-authored ones, suggesting increased efficiency but also raising concerns about the thoroughness of the review process.
        \item \agenticais can drastically scale developer output (e.g., the same developer producing 164 \agentprs in 3 days vs. 176 \humanprs over 3 years), yet these PRs alter fewer structural aspects of code (e.g., lower rates of cyclomatic complexity changes), raising new questions about the complexity and long-term maintainability of Agentic Code Changes.
    \end{itemize}
    \item A list of nine research directions that are derived from our empirical findings in \dataset{}, intended to guide future exploration at the intersection of SE and \agenticai.
\end{itemize}

The remainder of this paper is structured as follows. Section~\ref{sec:background} presents the background information. Section~\ref{sec:dataset} introduces the \dataset{} dataset collected in our study. Section~\ref{sec:applications} outlines three case studies, along with key findings and future research directions based on the \dataset{} dataset. Section~\ref{sec:discuss} reflects on the future of software engineering (SE 3.0) in the \agenticai era. Section~\ref{sec:conclusion} concludes the paper.

\section{Background}\label{sec:background}


\subsection{From SE 1.0 to SE 3.0}\label{sec:evolution}

Software engineering has undergone a profound evolution, marked by increasing system complexity and deeper integration of Artificial Intelligence (AI) into development workflows. This evolution can be understood through three major paradigms: SE 1.0, SE 2.0, and SE 3.0. Additionally, we propose an intermediate step—SE 1.5—to capture the transitional role of token-level AI assistance, as summarized in Table~\ref{tab:se_evolution}.

\textbf{SE 1.0 (Classical Software Engineering)} represented the traditional era, where software was hand-coded using explicit rules and deterministic flows (e.g., waterfall model). Programs were built and deployed in tightly controlled environments, requiring precise specifications and predictable behavior. No AI assistance was present.

\textbf{SE 1.5 (Predictive Coding)} introduced token-level, context-aware assistance such as autocomplete and smart suggestions. This paradigm includes tools that predict the next token, line, or code block a developer is likely to type, based on recent edits and the surrounding context. While useful for speeding up typing and reducing boilerplate, these tools operate only at a micro-scale and provide inline suggestions without autonomy. 

\vspace{-1mm}
\begin{itemize}
    \item \textit{Focus:} Micro-level suggestions during typing.
    \item \textit{Typical tools:} IntelliSense, IntelliCode, early versions of GitHub Copilot.
    \item \textit{Human role:} Choose or ignore suggestions in real-time.
\end{itemize}
\vspace{-1mm}

\textbf{SE 2.0 (AI-Assisted Software Engineering)} marks the advent of large language models (LLMs) in development environments. Tools in this category generate full functions, tests, or boilerplate code in response to natural-language prompts. These tools maintain file- or project-level context and assist with code explanation, translation, and review. Human developers remain in control, using the AI as a powerful assistant.

\vspace{-1mm}
\begin{itemize}
    \item \textit{Focus:} Macro-level generation and transformation of code.
    \item \textit{Typical tools:} GitHub Copilot, Amazon CodeWhisperer, Gemini Code Assist, Cursor, Windsurf.
    \item \textit{Human role:} Provide intent, review AI output, integrate or revise suggestions.
\end{itemize}
\vspace{-1mm}

\textbf{SE 3.0 (Agentic Software Engineering)} envisions software development as an intent-driven, conversational process where developers collaborate with autonomous AI teammates~\cite{hassan_se3_2024}. These agents operate at the task level—reading codebases, planning changes, executing tools, refactoring code, running tests, and submitting pull requests.

The developer’s role shifts to orchestration: setting goals, permissions, and reviewing final changes. These workflows typically extend beyond classic IDEs and into AI-native orchestration environments -- with recent experts projecting the death of the IDE by as early as end of 2025.

The role of Developers shifts from manual coding to high-level orchestration, setting goals, permissions, and reviewing completed work. Increasingly, workflows move beyond traditional IDEs into AI-native orchestration environments designed for human-agent collaboration. As Boris Cherny (lead engineer on Claude Code at Anthropic) observed, ``There’s a good chance that by the end of the year, people aren’t using IDEs anymore'' —suggesting that familiar development tools may give way to new agent orchestration platforms.

\vspace{-1mm}
\begin{itemize}
    \item \textit{Focus:} End-to-end execution of development tasks (design $\rightarrow$ code $\rightarrow$ test $\rightarrow$ commit).
    \item \textit{Typical tools:} Devin, GitHub Copilot Workspace, Google Jules, OpenAI Codex, Claude Code, Cursor Deep Agent, Genie.
    \item \textit{Human role:} Define goals, constraints, and permissions; review final changes; refine agent policies.
\end{itemize}
\vspace{-1mm}

\begin{table}[th]
\centering
\caption{Dimensions of AI Integration in Software Engineering}
\begin{tabular}{lccc}
\toprule
\textbf{Dimension} & \textbf{Predictive Coding (SE 1.5)} & \textbf{AI-Assisted SE (SE 2.0)} & \textbf{Agentic SE (SE 3.0)} \\
\midrule
Unit of work       & Tokens / lines                  & Functions / files              & Features / tickets \\
Autonomy           & None (inline hints)             & Low (single-shot generation)   & High (plan–execute–verify loops) \\
Context scope      & Current cursor position                & File / repo                    & Whole system + external tools \\
Primary benefit    & Typing speed                    & Rapid prototyping, refactoring & Scalable automation of dev tasks \\
Main risk          & Minor distraction               & Hallucinated code              & Unintended repo or infra changes \\
\bottomrule
\end{tabular}
\label{tab:se_evolution}
\end{table}


\subsection{\agenticai in Software Engineering}\label{sec:agentic}

\agenticais go far beyond predictive code completion. They can autonomously decompose complex goals, invoke external tools (e.g., debuggers, compilers, search engines), execute code, and iteratively refine their outputs with minimal human oversight. These capabilities mark a shift toward intelligent systems that exhibit decision-making, adaptation, and autonomy in practical development workflows. Early research prototypes illustrate this trend. For example, RepairAgent couples an LLM with a finite-state controller to automatically identify, edit, and test buggy code, successfully repairing real-world software defects with limited human input~\cite{bouzenia_repairagent_2025}. Cheng et al.~\cite{cheng2025brtagent} present BRT Agent, an LLM-powered system that autonomously generates bug reproduction tests (BRTs) from real bug reports at Google. These agent-generated BRTs improve automated program repair effectiveness, enabling more accurate fix selection using their proposed Ensemble Pass Rate (EPR) metric. These emerging systems share four defining properties of \agenticais: (i) persistent memory across dialogue turns and interactions, (ii) tool-use planning to select and sequence external operations, (iii) self-reflection to assess and revise intermediate results, and (iv) the ability to negotiate control by handing off tasks to human reviewers when needed. Together, these properties enable a new class of semi-autonomous intelligent systems that can serve as collaborators, not just assistants, in software engineering workflows.

\subsection{Benchmarking Autonomous Capabilities}\label{sec:benchmarks}

Evaluation of AI in software engineering has progressed from single-function code correctness to full-scale, project-level assessment. For example, SWE-bench evaluates whether models can resolve real GitHub issues in multi-file Python projects~\cite{jimenez2024swebench}. While early models showed low resolution rate, recent improvements highlight rapid progress. SWE‑bench Multimodal~\cite{yang2024swebenchmultimodal} extends these challenges to JavaScript with visual context. REPOCOD~\cite{liang2024repocod} evaluates repository-level code completion by combining retrieval of repository-level context with developer-written test cases. FeedbackEval~\cite{dai2025feedbackeval} investigates iterative, feedback-driven repair by assessing models' ability to use compiler and test feedback over multiple rounds. Together, these efforts reveal the gap between research-oriented benchmarks and real-world software engineering, highlighting the need for more comprehensive and practical evaluations.

\subsection{Human-AI Collaboration and Productivity}\label{sec:collaboration}

Empirical studies show that developers engage AI tools in two distinct modes: acceleration (speeding up familiar work) and exploration (probing unfamiliar design spaces)~\cite{forsgren_space_2021}.  
Productivity gains thus depend on balancing rapid code generation with the overhead of validation, testing, and trust calibration, dimensions captured by the multifaceted SPACE framework for developer productivity~\cite{forsgren_space_2021}.
Other research emphasizes the importance of human-AI interaction factors, particularly how developers form mental models of AI behavior and adapt through feedback. For example, Desolda et al.~~\cite{desolda2025understandingusermentalmodels} studied focus groups with 56 developers and introduced ATHENA, an adaptive code assistant designed to improve trust, usability, and integration in real-world workflows.
These findings suggest that as AI tools become more autonomous, successful integration will depend as much on aligning with human cognitive models and workflows as on raw technical capability.


\section{The \dataset{} Dataset}\label{sec:dataset}

In this section, we present the methodology used to construct the \dataset{} dataset, which captures the contributions of \agenticais to open-source software development on GitHub.

\subsection{Data Collection}

\begin{table}[t]
\centering
\caption{GitHub search query and collection start date for each \agenticai.}
\begin{tabular}{lll}
\toprule
 & GitHub Search Query & Start Date \\
\midrule
\codex & is:pr head:codex/ & 2025-05-16 \\
\devin & is:pr author:devin-ai-integration[bot] & 2024-12-24 \\
\copilot & is:pr head:copilot/ & 2025-01-01 \\
\cursor & is:pr head:cursor/ & 2025-01-01 \\
\claude & is:pr \enquote{Co-Authored-By: Claude} & 2025-02-24 \\
\bottomrule
\end{tabular}
\label{tab:pr_queries}
\end{table}

To construct the \dataset{} dataset, we mine pull requests (PRs) from GitHub that are associated with five widely used \agenticais: \codex, \devin, \copilot, \cursor, and \claude. These agents are selected based on their growing adoption and their explicit roles as autonomous contributors or co-authors in software repositories. We use the GitHub REST API\footnote{\url{https://docs.github.com/en/rest}} in combination with targeted search queries to extract PRs attributed to these agents. Table~\ref{tab:pr_queries} outlines the specific search queries used to identify PRs for each agent. These queries are adapted from PRarena\footnote{\url{https://github.com/aavetis/PRarena}} and leverage recognizable signals such as bot author names (e.g., \enquote{devin-ai-integration[bot]}), branch prefixes (e.g., \enquote{head:copilot/}), and explicit messages in the PR body (e.g., \enquote{Co-Authored-By: Claude}). We add a custom search query for \claude and, to ensure the quality of the results, apply a start date filter to exclude unrelated PRs. For \copilot, we additionally verify that the user login name is \enquote{copilot}, as this signal cannot be captured directly through the search query.

Release or public launch dates of \agenticais are determined from official announcements and documentation where available. For instance, \codex has clearly stated launch dates in its release announcement~\cite{openai_codex}. \devin\footnote{\url{https://docs.devin.ai/release-notes/overview\#december-24\%2C-2024}} and \claude\footnote{\url{https://docs.anthropic.com/en/release-notes/claude-code}} also have documented release dates in their respective release notes. For tools such as \cursor and \copilot, where no official release date was found, we adopted a conservative start date of January 1, 2025. The data collection cut-off date is June 22, 2025. Figure~\ref{fig:pr_cumulative_combined} visualizes the cumulative PR growth during this period, showing that the majority of PRs from \cursor and \claude occur after March 2025, which justifies the conservative start date of January 1, 2025.

\subsection{Dataset Overview}

\begin{table}[t]
  \centering
  \caption{Statistics of the \dataset{} dataset (full dataset) and \datasetpop{} (repositories with ${>}500$ stars).}
  \label{tab:datasets}
  \begin{subtable}[t]{0.48\linewidth}
    \centering
    \caption{\dataset{}: Full dataset.}
    \label{tab:dataset_full}
    \begin{tabular}{lrrr}
      \toprule
                    & \#PR & \#Developer & \#Repo \\ \midrule
      \codex        & 411,621 & 41,619 & 53,702 \\
      \devin        & 24,893  & 2,897  & 3,857 \\
      \copilot      & 16,531  & 1,916  & 3,097 \\
      \cursor       & 1,981   & 753    & 828 \\
      \claude       & 1,509   & 585    & 645 \\ \midrule
      \textbf{Total}& \textbf{456,535} & \textbf{47,303} & \textbf{61,453} \\
      \bottomrule
    \end{tabular}
  \end{subtable}
  \hfill
  \begin{subtable}[t]{0.48\linewidth}
    \centering
    \caption{\datasetpop{}: Filtered ($>$500 stars).}
    \label{tab:dataset_filter}
    \begin{tabular}{lrrr}
      \toprule
                    & \#PR & \#Developer & \#Repo \\ \midrule
      \codex        & 2,686 & 522 & 467 \\
      \devin        & 2,729 & 300 & 130 \\
      \copilot      & 1,462 & 309 & 215 \\
      \cursor       & 144   & 66  & 52 \\
      \claude       & 101   & 68  & 61 \\ \midrule
      \textbf{Total}& \textbf{7,122} & \textbf{1,240} & \textbf{856} \\
      \bottomrule
    \end{tabular}
  \end{subtable}
\end{table}

The collected \dataset{} dataset contains 456,535 pull requests opened by 47,303 distinct developers across 61,453 repositories (Table~\ref{tab:dataset_full}). Although \codex{} was released on 2025-05-16, it dominates the overall volume, surpassing \devin{}, the second highest, by approximately 17 times. To focus subsequent analyses on popular projects, we filter the \dataset{} dataset by retaining only repositories with $\ge\!500$ GitHub stars, resulting in a subset named \datasetpop{}. This filter reduces the sample to 7,122 PRs from 1,240 developers spanning 856 repositories (Table~\ref{tab:dataset_filter}), yet maintains representation from all five agents. Figure~\ref{fig:pr_cumulative_popular} shows the cumulative PR volume in \datasetpop{} .

For comparison, we stratified PRs from the same popular repositories created in 2025, excluding those created by \agenticais{}, to serve as human-created PRs, i.e., \textit{\humanprs}. This comparison set comprises 6,628 PRs that are created by 2,515 developers from 818 repositories~(we cannot sample enough PRs for some repositories). All downstream applications of the \dataset{} dataset presented in this paper (Section~\ref{sec:applications}) are based on \datasetpop{}.

\subsection{Dataset Schema}

\begin{figure}[t]
  \centering
  \includegraphics[width=\linewidth]{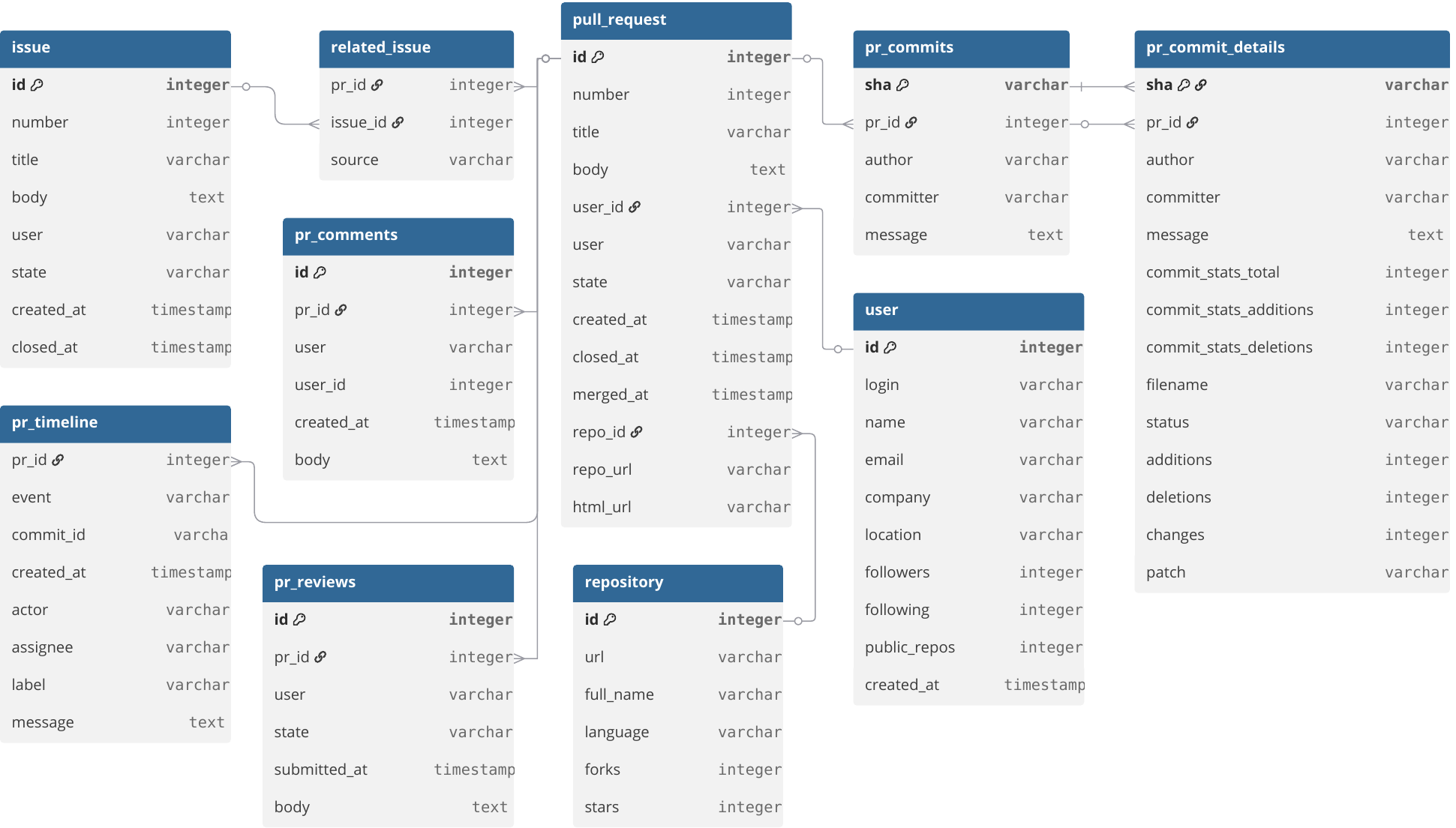}
  \caption{Dataset schema of the \dataset{} dataset.}
  \label{fig:dataset_schema}
\end{figure}

Figure~\ref{fig:dataset_schema} shows the dataset schema of the \dataset{} dataset. At its centre lies the \tablefmt{pull\_request} table. Each supporting table serves a distinct analytic purpose:

\begin{itemize}
  \item \tablefmt{repository}: records static project attributes such as name, primary language, star count, and fork count.
  \item \tablefmt{user}: stores profile attributes such as location, company affiliation, and follower count.
  \item \tablefmt{pr\_timeline}: tracks events for each PR—such as opened, assigned, labelled, merged—making it straightforward to reconstruct PR lifecycles.
  \item \tablefmt{pr\_reviews} and \tablefmt{pr\_comments}: capture feedback from reviewers and developers, along with timestamps and author identities, supporting studies of communication dynamics.
  \item \tablefmt{pr\_commits} and \tablefmt{pr\_commit\_details}: list the commit hashes included in each PR, and provide file-level diff statistics (added, deleted, modified) and the raw patch, supporting commit-level studies (e.g., code changes).
  \item \tablefmt{related\_issue} and \tablefmt{issue}: link PRs to issues they close or reference, and provide issue metadata, enabling tracing back to the source event that triggered the PR (e.g., bug reports or feature requests).
\end{itemize}

All tables are provided as CSV files, accompanied by scripts that recreate the dataset schema. The dataset is included in our replication packages along with example Python notebooks to minimize onboarding friction for future researchers.

\section{Applications of the \dataset{} dataset}~\label{sec:applications}

In this section, we demonstrate the applications of our \dataset{} dataset by conducting three case studies. We highlight our insights from the \dataset{} dataset, including ten findings and nine research directions. 



\subsection{Productivity in \agenticais Era}\label{sec:productivity}

Understanding how \agenticais affect developer productivity at scale is essential before introducing new solutions into already complex workflows. With a rapidly expanding landscape of \agenticais such as \claude, \codex, and others, it is tempting to adopt the latest innovations even without empirical evidence of their effectiveness. However, adopting new tools risks adding noise, becoming counterproductive rather than valuable. Instead, we propose to take a more disciplined approach for software engineering research: carefully evaluating which \agenticais meaningfully support real-world development through rigorous measurements. Only by measuring the actual impact of these emerging \agenticais can we identify which tools genuinely enhance productivity and warrant prioritization.

To move towards this goal, our first effort focuses on systematically categorizing software engineering tasks within a representative set of \agentprs. Specifically, we construct \datasetpop{}, a filtered high-quality dataset of PRs from popular open-source repositories (each with over 500 GitHub stars as of June 22, 2025). Each PR (title and body) is automatically classified into one of 11 task categories defined by Conventional Commits Specification\footnote{\url{https://www.conventionalcommits.org}} (e.g., \texttt{feat}, \texttt{fix}, \texttt{docs}) using \textit{GPT-4.1-mini}. Such state-of-the-art LLMs have demonstrated strong performance in annotating software engineering artifacts~\cite{toufique2025llmlabel}, making them suitable for this task. This categorization provides a structured view of how \agenticais are being used across different software development activities.

Building on this categorization, we evaluate two key indicators that reflect the quality and efficiency of \agentprs contribution. First, we compute the acceptance rate, i.e., the proportion of \agentprs that are successfully merged. We calculate the acceptance rate per task category for each \agenticai, and compare with \humanprs. Second, for agents that provide sufficient metadata, we analyze during-work time~(i.e., resolution time)~\cite{nastos2025interpretable} to estimate how long \agenticai actively worked on each PR. For example, in the case of \copilot, we measure the elapsed time from PR creation to the point when the agent finishes its job~(distinct from the time until the PR is merged or closed). These metrics enable a quantitative comparison between agent- and human-authored contributions, offering actionable insights into which \agenticais are most effective in real-world development settings.

%
%

Next, we present our experimental findings and discuss their implications in depth:

\begin{table}[tbp]
\caption{Percent distribution of PR task types for \agenticais in \datasetpop{}, along with the total number of PRs.}
\label{tab:pr_task_type}
\begin{tabular}{@{}lrrrrrrrrrrrr@{}}
\toprule
 & \%feat & \%fix & \%perf & \%refac & \%sty & \%doc & \%test & \%chore & \%build & \%ci & \%other & \#total \\
\midrule
Human & \textbf{29.4} & 26.9 & 1.3 & 5.6 & 0.9 & 7.8 & 2.8 & 12.8 & 9.3 & 1.5 & 1.7 & 6,618 \\ \midrule
\codex & \textbf{28.4} & 27.2 & 0.6 & 7.1 & 0.8 & 21.3 & 9.2 & 2.6 & 1.6 & 1.1 & 0.1 & 2,686 \\
\devin & \textbf{35.8} & 27.4 & 1.6 & 10.4 & 0.4 & 12.0 & 3.0 & 6.2 & 1.9 & 1.0 & 0.1 & 2,729 \\
\copilot & 29.7 & \textbf{42.2} & 1.4 & 5.2 & 0.3 & 9.8 & 4.4 & 1.8 & 3.5 & 1.4 & 0.2 & 1,462 \\
\cursor & \textbf{41.7} & 34.7 & 0.7 & 2.8 & 1.4 & 6.9 & 2.8 & 4.2 & 3.5 & 0.0 & 1.4 & 144 \\
\claude & \textbf{49.5} & 18.8 & 0.0 & 8.9 & 0.0 & 13.9 & 3.0 & 1.0 & 3.0 & 2.0 & 0.0 & 101 \\
\bottomrule
\end{tabular}
\end{table}


\finding{\agenticais have moved beyond the role of mere coding assistants, becoming indispensable collaborators, as recent 2025 trends suggest, each contributing unique strengths across a wide range of software engineering tasks such as feature development and bug fixing.} 
Table~\ref{tab:pr_task_type} shows that over 55\% of PRs submitted by \agenticais are either feature development (\texttt{feat}) or bug remediation (\texttt{fix}), matching the overall human distribution. However, individual agents diverge in how they allocate effort: \codex and \devin exhibit a well-balanced mix of various SE capabilities, closely mirroring human developers' task profiles. By contrast, 42.2\% of the \copilot Agentic-PRs are bug fixing PRS, surpassing both human developers and other agents. This emphasis is likely driven by \copilot's tight integration with GitHub issues, which frequently assigns it targeted bug-related tasks. Finally, \cursor and \claude appear to be primarily used for feature implementation, each submitting over 40\% of their PRs as features. However, since these two tools are mainly used locally, their relatively low usage, as indicated by PR counts, suggests that their characteristics would be better understood through analysis within local environments rather than relying solely on GitHub data.

\begin{figure}[h]
  \centering
  \includegraphics[width=\linewidth]{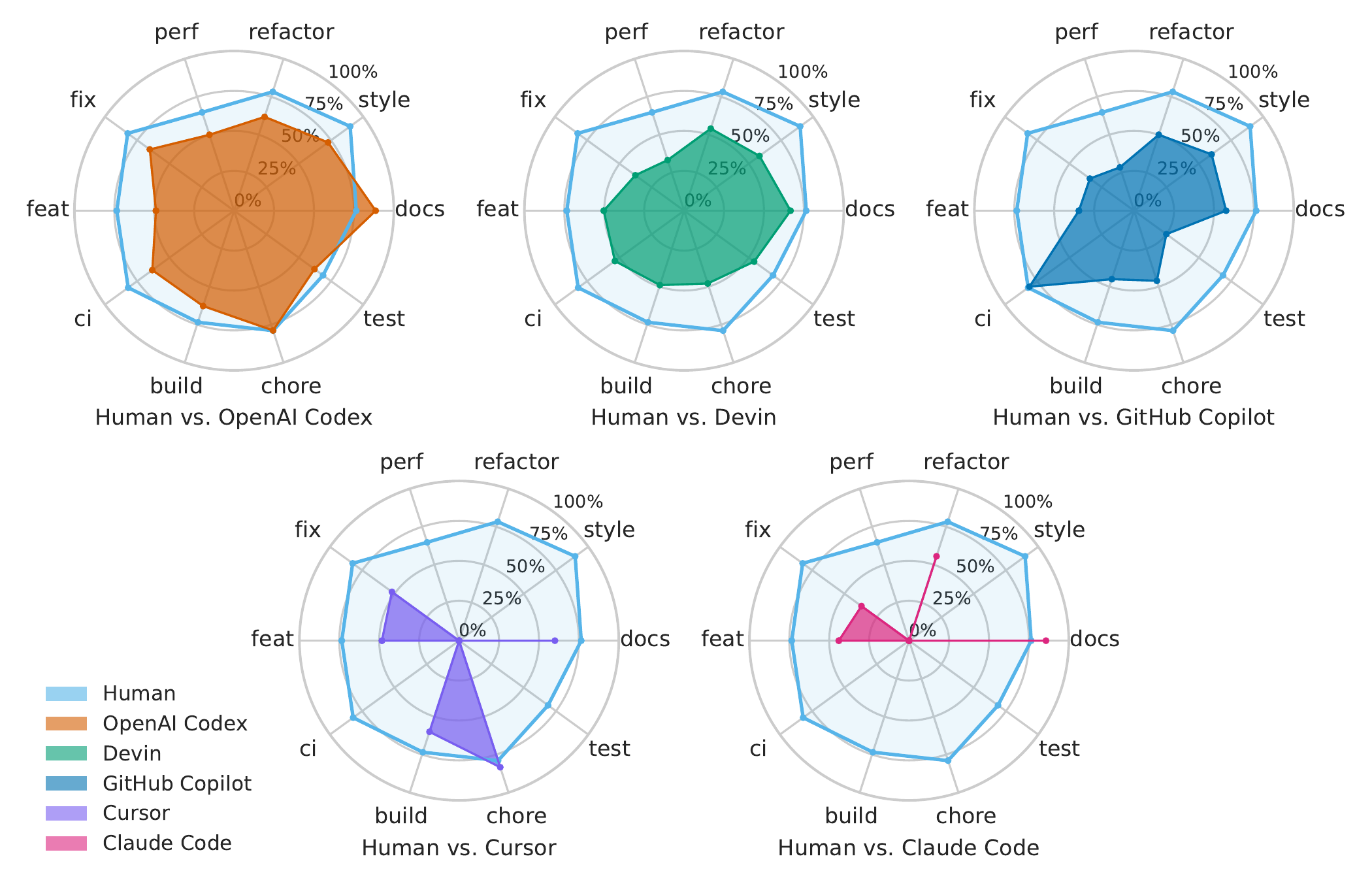}
  \caption{PR acceptance rate by task category for Human vs. \agenticais (each subplot compares one \agenticai against the human baseline).}
  \label{fig:pr_merge_compare_radar}
\end{figure}

\finding{\agenticais lag human in PR acceptance rates by a large margin, especially on complex tasks such as feature development and bug fixing.} 
Figure~\ref{fig:pr_merge_compare_radar} reveals a striking pattern: across all evaluated \agenticais, PR acceptance rates consistently lag behind human performance, particularly for feature development (\texttt{feat}), bug fixes (\texttt{fix}), and performance optimization (\texttt{perf}) tasks. Among the \agenticais, \codex achieves the highest acceptance rate at 64\%, followed by \devin at 49\% and \copilot at 35\%. These substantial performance gaps, ranging from 15 to 40 percentage points below human performance, suggest systemic limitations rather than isolated implementation flaws. Notably, these results contrast sharply with benchmark performance on SWE-bench verified~\cite{jimenez2024swebench}, a human-validated benchmark, where top-performing AI solutions report success rates exceeding 70\%\footnote{\url{https://www.swebench.com}}. This significant disparity between benchmark and real-world performance raises important questions about the ecological validity of current evaluation methodologies for \agenticais in software engineering contexts. These concerns about real-world AI performance are further supported by recent empirical evidence showing that AI tools actually increased task completion time by 19\% among experienced developers, despite expectations of productivity gains~\cite{metrMeasuringImpact}.

We do note that these performance gaps should not be interpreted as a sign of failure; they should be seen as a testament to how early we are in this transformation. These results were captured just two months into the public release of many of these agents. The fact that autonomous AI teammates are already operating across tens of thousands of repositories, submitting hundreds of thousands of PRs, and successfully landing complex changes—even at lower rates—is itself astonishing. As tools, RL training paradigms, frameworks evolve and best practices emerge, these early shortfalls will prove to be the launchpad for rapid and sustained improvement. The revolution is already underway and it is moving fast.

\finding{Documentation emerges as a clear strength of \agenticais.} 
In Figure~\ref{fig:pr_merge_compare_radar}, both \codex (88.6\%) and \claude (85.7\%) outperform the human baseline (76.5\%) in documentation-related PRs. Since documentation tasks primarily involve natural language generation rather than complex program logic, such tasks align well with the core strengths of LLM-powered \agenticais. The high acceptance rates for documentation-related PRs may also reflect greater reviewer confidence in AI-generated documentation. Unlike code changes, documentation modifications pose minimal functional risk, making documentation tasks a promising area for \agenticais applications.

\begin{figure}[tbp]
  \centering
  \includegraphics[width=0.7\linewidth]{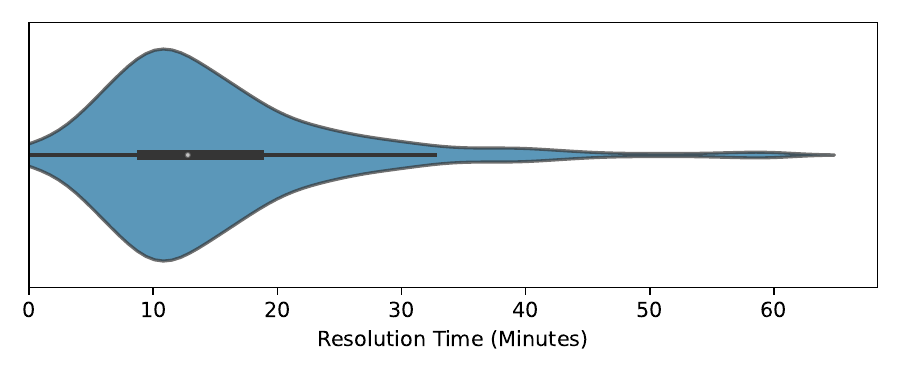}
  \caption{Distribution of \copilot resolution time.}
  \label{fig:copilot_job_completion_time}
\end{figure}

\finding{\copilot completes 75\% of its PR jobs in under 18.5 minutes, faster than typical human issue resolution, which usually takes more than half a day.} 
As shown in Figure~\ref{fig:copilot_job_completion_time}, \copilot delivers half of its PRs within 12.8 minutes. While 75\% of jobs are completed within 18.5 minutes, the distribution exhibits a long tail extending up to 60 minutes, with the 95th percentile exceeding one hour. Despite strong average performance, this tail latency suggests potential infrastructure issues such as cold starts, dependency resolution, or API throttling. In human-AI collaboration, such variability may hinder real-time feedback loops and delay CI/CD pipelines. In contrast, analyses of Apache Jira projects show that human developers typically resolve issues over multi-day periods. More than 50\% of Major and Critical issues take over 5 days, and even Trivial and Minor issues often~(more than 50\%) take over half a day~\cite{nastos2025interpretable}. These disparities underscore \copilot's advantage in job completion speed, enabling tighter iteration cycles and improving CI/CD responsiveness.



\researchdir{Develop integration-oriented benchmarks grounded in real-world software engineering workflows.} Our findings expose a key limitation reflecting AI's applications to software engineering: models generate syntactically valid code but often fail the nuanced human judgments essential for real-world acceptance. Unlike synthetic benchmarks that measure curated datasets~\cite{chen2021evaluating,jimenez2024swebench}, actual PR merges depend on factors such as maintainability, style, and project-specific context, elements that are central to practical adoption. Our \dataset{} dataset captures these workflows, enabling future benchmarks that evaluate what truly matters: not just functional correctness, but seamless integration into developer practices. This shifts the focus from ``Can we leverage AI to generate code?'' to ``Do AI contributions improve real-world software development?'' This marks a vital step towards deeper and closer evaluation of AI's actual impact on everyday software engineering workflows. It also helps identify where current \agenticais fall short compared to human developers, specially in terms of \textit{atomic} software engineering capabilities, in turn guiding targeted improvements.
%


\researchdir{Analyze rejected PRs to identify \agenticai failure modes.} Rejected PRs often contain valuable diagnostic artifacts---such as review comments, timeline events, and patch-level feedback---that are preserved in the \dataset{} dataset. By mapping GitHub events to specific failure modes, including logic bugs, syntax errors, inadequate testing, or misalignment with project conventions, researchers can classify errors and develop targeted mitigation strategies. This analysis complements static benchmarks and informs the design of agents with retry capabilities or self-diagnostic features, ultimately enabling reflexive agents that can detect and address their own mistakes.
%

\researchdir{Latency-aware orchestration for \agenticais.} The heavy-tailed latency we observe in \copilot's traces reveals systemic bottlenecks that undermine the timely feedback cycles expected in modern CI/CD workflows. Rich telemetry recorded in GitHub Actions logs\footnote{\url{https://github.com/NethermindEth/nethermind/actions/runs/15811046373/job/44562368029}} already captures environment setup, MCP server start-up, and inference steps of \copilot, providing a concrete foundation for root-cause analysis. Researchers can therefore investigate systems-level remedies such as caching, speculative execution, incremental previews, or warm-started runtime systems. In addition, prompt-level efficiency techniques merit further study. For instance, the Chain-of-Draft prompting strategy reduces token usage to approximately 55\% of Chain-of-Thought while retaining over 90\% of solution quality on SWE-bench, resulting in roughly 45\% shorter wall-clock times for code-fix tasks~\cite{yang2025chainofdraft}. Combining such token-efficient reasoning with adaptive workload orchestration has the potential to reduce both median and tail latencies, helping \agenticais better deliver results within a possibly shorter and more predictable timeframe rather than occasionally leaving users waiting indefinitely.

\subsection{Agentic Code Review Dynamics}\label{sec:review}


Section~\ref{sec:productivity} shows that \agenticai activities on GitHub are increasing rapidly, with such agents generating PRs at scale. This trend exposes a critical yet often overlooked bottleneck: despite rapid advancements in code generation capabilities and volume, human-centric code review processes have remained largely unchanged. Consequently, the scalability of \agenticai contributions has created a growing imbalance between PR submission rates and review capacity. Traditional software engineering workflows already experience possible delays in feedback and integration, which are further exacerbated by agentic workflows that accelerate contribution rates beyond the capacity of existing review mechanisms, thereby risking bottlenecks in quality assurance and integration/release cycle. In this section, we examine how review dynamics are adapting to this evolving context by answering two key questions: (1) how quickly agentic contributions are processed, and (2) who is responsible for reviewing them.

For each PR, we calculate its review turnaround time, defined as the elapsed duration from PR creation to closure (either acceptance or rejection), to compare review efficiency between \agentprs and \humanprs. To assess whether differences in turnaround times are statistically significant, we perform the Mann-Whitney U test~\cite{Mann1947OnAT} at a significance level of $\alpha=0.05$, conducted separately for accepted and rejected PRs. Additionally, we quantify effect sizes using Cliff's delta~$d$~\cite{Cliff}, interpreting results according to established thresholds from prior studies~\cite{Cliff_threshold}.


In addition, we extract PRs and their corresponding reviews from \datasetpop{}. Reviewer identities are classified as \textit{human} or \textit{bot} based on username patterns (e.g., the suffix \texttt{[bot]}) and GitHub's \texttt{type} field. Each PR is then categorized into one of four groups: reviewed exclusively by humans, exclusively by bots, by both, or by neither. We further identify the 10 most active bot reviewers among \agentprs and analyze their review activities in comparison to human reviewers.


\begin{figure}[t]
  \centering
  \includegraphics[width=0.9\linewidth]{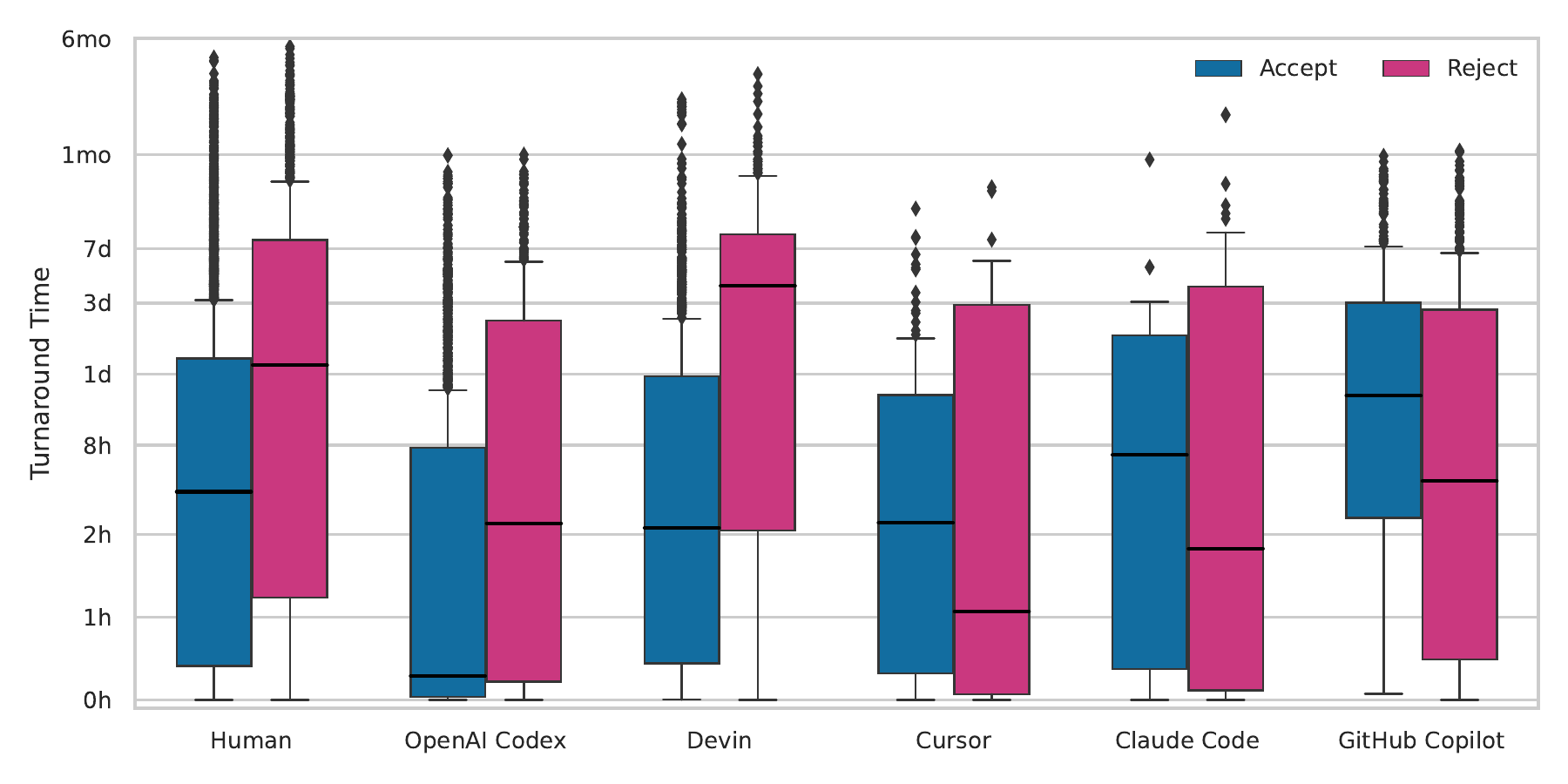}
  \caption{Distribution of turnaround times for accepted and rejected PRs across agents and humans.}
  \label{fig:turnaround_distribution}
\end{figure}

\begin{table}[t]
\caption{Turnaround time comparison for \agentprs and \humanprs. MT.: median turnaround time (in hours); Sig.: statistical significance of difference (\cmark = $p<0.05$).}
\label{tab:turnaround_statistical}
\begin{tabular}{lrrrrlrrrr}
\toprule
             & \multicolumn{4}{c}{Accepted PRs}                  &  & \multicolumn{4}{c}{Rejected PRs}                  \\ \cmidrule(lr){2-5} \cmidrule(l){7-10} 
             & \%Accepted & MT. (h) & Sig. & Effect Size    &  & \%Rejected & MT. (h) & Sig. & Effect Size    \\ \midrule
Human        & 76.8 & 3.9               & NA    &  NA              &  & 16.1 & 27.6              &   NA  &     NA           \\ \midrule
\codex & 65.3 & 0.3               & \cmark   & medium ($+$)     &  & 21.3 & 2.4               & \cmark   & small ($+$)      \\
\devin        & 48.9 & 2.2               & \cmark   & negligible ($+$) &  & 48.4 & 93.9              & \cmark   & negligible ($-$) \\
\copilot      & 38.2 & 17.2              & \cmark   & small ($-$)      &  & 32.1 & 4.6               & \cmark   & small ($+$)      \\
\cursor       & 51.4 & 2.4               & \xmark   & NA              &  & 21.5 & 1.1               & \cmark   & small ($+$)      \\
\claude  & 52.5 & 6.9               & \xmark   & NA              &  & 32.7 & 1.8               & \cmark   & small ($+$)      \\ \bottomrule
\end{tabular}
\end{table}

\finding{OpenAI Codex PRs cut review time by 10 times, boosting efficiency but questioning review depth.}
As shown in Figure~\ref{fig:turnaround_distribution} and Table~\ref{tab:turnaround_statistical}, accepted PRs from \codex close in a median of 0.3 hours~(18 minutes), which is significantly faster than \humanprs~(3.9 hours) with a medium effect size. Rejected PRs from \codex are also triaged significantly faster (2.4 vs. 27.6 hours) with a small effect size. This suggests that \codex PRs are often either unambiguously correct or noticeably flawed, allowing reviewers to make rapid decisions.
In contrast, \copilot shows the opposite pattern: slow to accept, fast to reject. Accepted \copilot PRs take 17.2 hours to review, which is over 4 times longer than accepted \humanprs (3.9 hours), with a small negative effect size. In contrast, its rejected PRs are reviewed in just 4.6 hours, significantly faster than human rejections (27.6 hours). This disparity implies that reviewers struggle more with validating or iterating \copilot's PRs, whereas flawed ones are rapidly dismissed. This may reflect quality inconsistency or ambiguity in \copilot outputs that demand closer scrutiny.

In addition, rejected PRs from \agenticais are generally triaged faster than \humanprs, saving reviewer time despite higher rejection rates. Table~\ref{tab:turnaround_statistical} shows that across all \agenticais except \devin, rejected \agentprs are closed significantly faster than those from humans. While these trends suggest improved efficiency, they raise concerns about review depth. Given the scale of contributions and the lack of strong accountability mechanisms, this rapid pace may suggest superficial review rather than genuine confidence. Without assessing the presence and depth of review comments or formal, actionable change requests, it remains uncertain whether these Agetnic-PRs undergo meaningful scrutiny or are merely triaged.



\begin{figure}[tbp]
  \centering
  \includegraphics[width=0.8\linewidth]{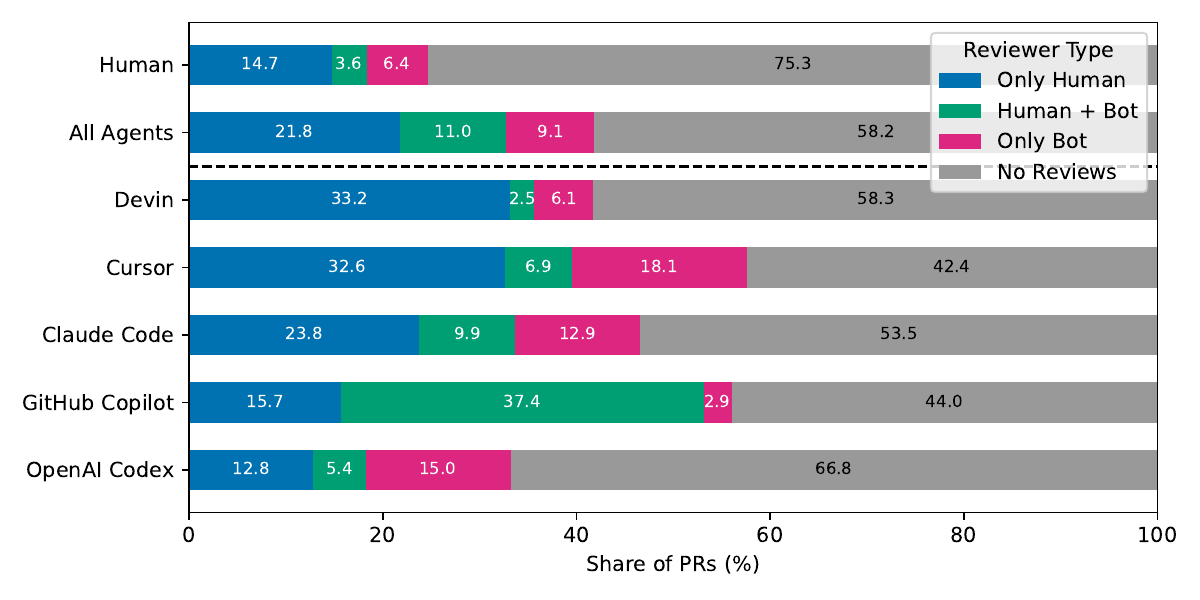}
  \caption{Reviewer types across the PRs in \datasetpop{}.}
  \label{fig:reviewer_type}
\vspace{-3mm}
\end{figure}

\finding{Human reviewers remain dominant across \agentprs yet \copilot drives a shift toward automated hybrid collaboration in review.} 
Figure~\ref{fig:reviewer_type} shows that both \humanprs and \agentprs receive no explicit review in the majority of cases~(75.3\% and 58.2\%, respectively), while the second most common category involves reviews conducted solely by humans, at 14.7\% and 21.8\%. Notably, bot reviewers are more prevalent in \agentprs~(20.1\%) than in \humanprs (10.0\%).
Breaking down review dynamics by tool reveals distinct patterns. In \devin, \cursor, and \claude, human reviewers remain dominant, with 33.2\%, 32.6\%, and 23.8\% of PRs reviewed solely by humans, respectively. \codex, by contrast, presents a more balanced distribution, with bot-only reviewers roughly matching with human-only reviewers. Most notably, \copilot shows a marked shift toward a hybrid human-bot review mode, where 37.4\% of PRs involve both. In such cases, bots commonly function as co-reviewers\footnote{\url{https://github.com/kanisterio/kanister/pull/3505}} or as first-pass screeners,\footnote{\url{https://github.com/ryoppippi/ccusage/pull/128}} providing initial assessments that can benefit the subsequent human review. This trend highlights a potential transformation in the allocation of review responsibilities in \copilot and may significantly reduce the need for manual triage if this pattern continues to grow.

\begin{figure}[t]
  \centering
  \includegraphics[width=0.7\linewidth]{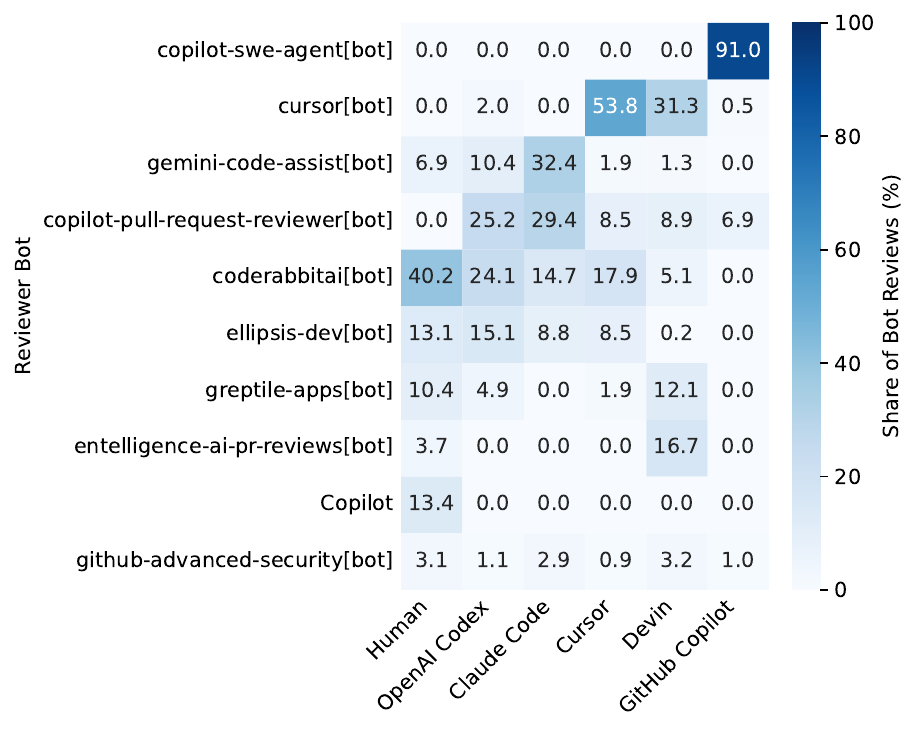}
  \caption{Top-10 bot reviewers and their review distribution across PRs from different agents.}
  \label{fig:bot_heatmap}
\vspace{-3mm}
\end{figure}

\finding{\agenticais and their paired review bots often originate from the same provider, forming closed PR-review loops that streamline workflows; but risk reinforcing provider-specific biases.}
We analyze the top 10 most active review bots to understand on which \agenticais they operate. Figure~\ref{fig:bot_heatmap} illustrates a strong association between review bots and \agenticais from the same provider. For example, PRs from \copilot are often reviewed by \texttt{copilot-pull-request-reviewer[bot]}, responsible for 91.0\% of bot reviews on \copilot PRs. Similarly, \texttt{cursor[bot]} is responsible for reviewing 13.5\%~(53.8\% of bot reviews) of PRs submitted through \cursor. 

These provider-aligned PR-review loops offer practical advantages with developers benefitting from tightly integrated workflows, consistent UX, and faster iteration cycles. However, they also raise concerns about internal bias: when a review bot repeatedly evaluates code produced by an \agenticai from the same provider, it may overlook stylistic issues or latent errors that fall outside its tuned assumptions. In contrast, bots like \texttt{coderabbitai[bot]} operate across both human- and agent-authored code and are not confined to a single provider ecosystem. Their broader exposure may enable more robust and diverse code evaluation.

Finally, among the top 10 review bots, only \texttt{github-advanced-security[bot]} remains rule-based rather than AI-powered. Its continued presence highlights a notable frontier—security. For critical tasks like vulnerability detection, traditional static analysis tools still outperform current AI systems. The absence of trusted AI-powered security reviewers underscores an important research direction: designing \agenticais that can operate reliably and transparently in safety-critical domains.


\researchdir{Understand and reduce the human cost of reviewing agentic code changes.} The cost of reviewing \agentprs lies not only in their correctness, but in the ambiguity and required mental effort to validate them. Although some \agenticai submitted \agentprs are reviewed rapidly (Figure~\ref{fig:turnaround_distribution}), it remains unclear whether this indicates reviewer confidence, superficial engagement, or a lack of oversight. Future research could leverage the \dataset{} dataset to capture signals of review effort, such as review comment volume, presence of requested changes, follow-up commits addressing feedback, or the time taken to integrate any received feedback. These signals could help researchers design new agent capabilities or workflows aimed at reducing review burden. For example, \agenticais may explain their code changes, proactively resolve merge conflicts, or minimize diff complexity. Our \dataset{} dataset supports such potential investigations with detailed PR timelines, review histories, and reviewer metadata. This research aligns with the growing research interest in human-aligned AI behaviour, and opens new directions for making autonomous contributions more interpretable, reviewable, and trustworthy.


\researchdir{Understand and improve how agentic code review contributes to the quality and process of pull request workflows.} As review bots such as \botfmt{coderabbitai} and \botfmt{copilot-pull-request-reviewer} become a routine part of software development workflows~(Figure~\ref{fig:bot_heatmap}), and \agenticais become more active in writing and changing code, the importance of code review continues to grow. It remains a key step to ensure software quality, especially when more and more code will be produced by AI. However, current evaluations of automated code review tools rely on developer opinions or post-hoc metrics~\cite{cihan2024automatedcodereviewpractice}, which do not fully answer questions such as how useful these reviews are. To help move the field forward, our dataset can be leveraged to study review behaviour in more detail, especially in PRs involving \agenticais. Review histories can be analyzed, allowing future research to investigate the kinds of feedback that review bots give, how such feedback connects to whether a PR gets merged, and how bots work alongside human reviewers. Especially as human-AI collaboration becomes more common, our dataset offers a solid starting point for future research that wishes to understand, evaluate, and improve how agentic code review fits into real-world workflows, thus improving software quality.

\researchdir{Design review triage systems that manage reviewer effort in response to the scale and complexity of AI-generated contributions.} As \agenticais generate PRs at scale, the allocation of human reviewer effort quickly becomes a bottleneck. To manage this increasing review workload, triage systems must go beyond static rules and incorporate dynamic factors such as PR complexity, risk level, and reviewer availability. For example, low-risk PRs like documentation edits can be routed to junior developers or bots, while high-impact system-level changes should be escalated to experienced reviewers. Research at Meta has shown that reviewer recommender systems, when tuned to account for reviewer workload and response latency, can reduce review times without sacrificing quality~\cite{rigby2025review}. Our \dataset{} dataset includes metadata such as reviewer identities, timestamps, and task categories, which makes it possible to identify and evaluate triage strategies. Future work could use this dataset to measure how different triage approaches impact metrics such as reviewer time saved, workload distribution, and the reduction of review smells. A deeper understanding of agentic review processes is essential to scale human oversight in workflows increasingly shaped by AI. As PR volume continues to grow, effective triage mechanisms will be essential to maintain reviewer capacity without compromising review quality or speed.

\subsection{Beyond Agentic/Vibe Coding for Software Engineering}\label{sec:coding}

The integration of \agenticais into software development workflows is rapidly accelerating, as reflected by the over 456K GitHub PRs now containing AI-generated code. These agents are no longer just assisting with isolated tasks, as they now actively shape development practices ranging from individual programming to team-based collaboration on GitHub. This growing presence of AI agents in the software lifecycle presents both significant opportunities and pressing questions about its long-term impact on software quality.

While AI tools can dramatically increase output, raw quantity is no longer the primary bottleneck. The core concern is whether AI-generated code meets the standards of correctness, maintainability, and robustness required in production environments. To better understand this, we analyze over 7K \agentprs drawn from 856 repositories with more than 500 stars, using metrics like cyclomatic complexity to evaluate code structure and potential maintainability costs. However, measuring code quality alone is insufficient. Without clear attribution, it is difficult to assign responsibility or intervene when issues arise. Moreover, AI behaviour is not monolithic and varies by programming language, domain, and workflow. We will explore the above aspects to understand how to guide the responsible integration of \agenticais into software engineering.

%

\begin{figure}[t]
  \centering
  \includegraphics[width=0.75\linewidth]{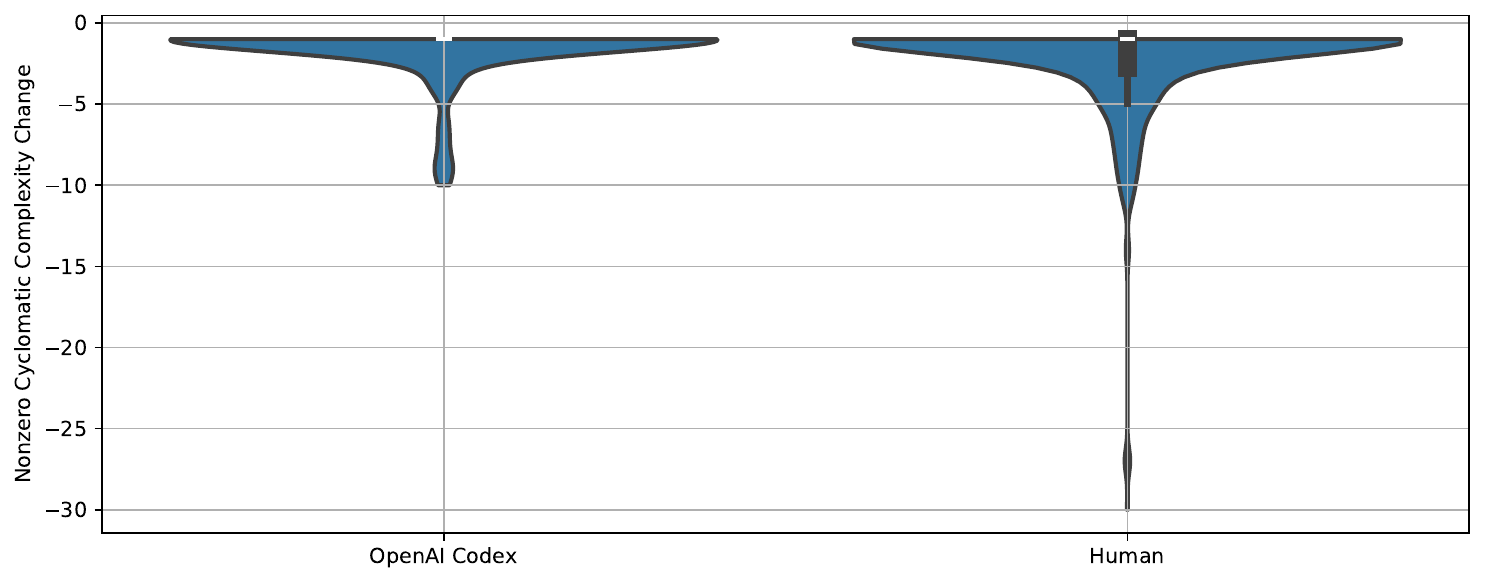}
  \caption{Distribution of non-zero cyclomatic complexity change by \codex vs. human.}
  \label{fig:compare_cc}
\vspace{-3mm}
\end{figure}

\finding{AI-assisted coding prioritizes throughput over complexity.} Our analysis of \codex usage, using the OpenHFT/Chronicle-Wire project as a case study, reveals an notable shift in development dynamics. Between May 30 and June 3, 2025, a single developer submitted 164 Codex-assisted PRs, nearly matching the 176 human-authored PRs produced over the preceding three and a half years (from November 2021 to May 2025). Despite this surge in output, only 9.1\% of the \agentprs introduced changes in cyclomatic complexity, compared to 23.3\% for \humanprs. Figure~\ref{fig:compare_cc} shows the distribution of non-zero complexity changes, suggesting that agentic code changes tend to favour minimal, boilerplate-style updates over more intricate implementations. This tradeoff, i.e., sheer throughput at the expense of per-PR intricacy, raises key questions: Does reduced complexity lead to fewer defects or faster review? Or, conversely, might this simplicity be a liability in domains where complexity, such as in highly optimized code, is essential?

\finding{Clear authorship attribution remains limited but essential for accountability.} Among the \agentprs, we found that \devin, \copilot, and \cursor explicitly indicate their authorship in commit metadata. In contrast, \codex and \claude attribute their contributions to human developers. While \claude includes a default \enquote{Co-Authored-By: Claude} message (which can be disabled), \codex provides no attribution at all. This lack of authorship traceability undermines core principles of responsible AI—such as transparency, auditability, and accountability—especially in the context of software engineering. For instance, tracing precise authorship in Codex-assisted PRs is often difficult due to incomplete or obscured metadata, even among follow-up commits.\footnote{\url{https://github.com/stanfordnlp/dspy/pull/8319}} Without clear ownership, it becomes harder to perform post hoc debugging, triage bugs, or assign responsibility for regressions. We argue that standardized authorship labelling should be a fundamental requirement for all AI-assisted development practices. Clear attribution not only enables code provenance tracking and behavioural analysis of \agenticais, but also supports more robust human-AI collaboration and aligns with emerging norms in responsible AI governance. Moreover, linking ownership-labelled code to artifacts like issue reports and bug fixes would open new opportunities to study how AI- versus human-authored code affects software quality and maintainability over time.



\finding{\agenticais exhibit distinct language preferences reflecting domain specialization in their capabilities.}
To examine how different \agenticais integrate into development workflows, we analyze the programming languages used in repositories where their PRs are submitted. Table~\ref{tab:language_usage} shows that TypeScript is the most common language across all agents, underscoring its popularity in AI-assisted development. However, notable divergences emerge: \codex shows a pronounced skew toward Python, while \copilot heavily favours C\#, likely reflecting their respective integrations and user bases. These variations suggest that each agent has developed domain-specific affinities, shaped by tooling ecosystems, developer intent, and integration contexts. Understanding these language-specific patterns is a key step toward evaluating agent behavior with greater precision and building agents better tuned to diverse development environments.

\begin{table}
\caption{Top-10 languages used in GitHub repositories across agents in \datasetpop{}.}
\label{tab:language_usage}
\begin{tabular}{lrrrrrrrrrrrrrrlrr}
\toprule
 &
  \multicolumn{2}{l}{All Agents} &
  \multicolumn{1}{l}{} &
  \multicolumn{2}{c}{OpenAI Codex} &
  \multicolumn{1}{l}{} &
  \multicolumn{2}{c}{Devin} &
  \multicolumn{1}{l}{} &
  \multicolumn{2}{c}{GitHub Copilot} &
  \multicolumn{1}{l}{} &
  \multicolumn{2}{c}{Cursor} &
   &
  \multicolumn{2}{c}{Claude Code} \\ \cmidrule(lr){2-3} \cmidrule(lr){5-6} \cmidrule(lr){8-9} \cmidrule(lr){11-12} \cmidrule(lr){14-15} \cmidrule(l){17-18} 
 &
  \# &
  \% &
  \multicolumn{1}{l}{} &
  \# &
  \% &
  \multicolumn{1}{l}{} &
  \# &
  \% &
  \multicolumn{1}{l}{} &
  \# &
  \% &
  \multicolumn{1}{l}{} &
  \# &
  \% &
   &
  \# &
  \% \\ \midrule
TypeScript       & \textbf{226}            & \textbf{26.4}                 &  & 117 & 25.1 &  & \textbf{71} & \textbf{54.6} &  & 36 & 16.7 &  & \textbf{24} & \textbf{46.2} &  & \textbf{16} & \textbf{26.2} \\
Python           & 172            & 20.1                 &  & \textbf{119} & \textbf{25.5} &  & 24 & 18.5 &  & 20 & 9.3  &  & 12 & 23.1 &  & 14 & 23.0 \\
C\#               & 78             & 9.1                  &  & 12  & 2.6  &  & 1  & 0.8  &  & \textbf{64} & \textbf{29.8} &  & 1  & 1.9  &  & 1  & 1.6  \\
Go               & 72             & 8.4                  &  & 37  & 7.9  &  & 10 & 7.7  &  & 18 & 8.4  &  & 3  & 5.8  &  & 9  & 14.8 \\
Rust             & 49             & 5.7                  &  & 28  & 6.0  &  & 10 & 7.7  &  & 6  & 2.8  &  & 5  & 9.6  &  & 4  & 6.6  \\
C++              & 42             & 4.9                  &  & 25  & 5.4  &  & 0  & 0.0  &  & 13 & 6.0  &  & 1  & 1.9  &  & 3  & 4.9  \\
JavaScript       & 40             & 4.7                  &  & 23  & 4.9  &  & 4  & 3.1  &  & 10 & 4.7  &  & 0  & 0.0  &  & 3  & 4.9  \\
Java             & 25             & 2.9                  &  & 18  & 3.9  &  & 0  & 0.0  &  & 6  & 2.8  &  & 0  & 0.0  &  & 1  & 1.6  \\
C                & 20             & 2.3                  &  & 10  & 2.1  &  & 3  & 2.3  &  & 6  & 2.8  &  & 1  & 1.9  &  & 0  & 0.0  \\
PHP              & 18             & 2.1                  &  & 12  & 2.6  &  & 1  & 0.8  &  & 1  & 0.5  &  & 3  & 5.8  &  & 2  & 3.3  \\ \bottomrule
\end{tabular}
\vspace{-3mm}
\end{table}

\researchdir{Understand and evaluate the long-term quality of agentic code contributions.} While our findings suggest that AI-generated code may simplify certain maintenance tasks, critical questions remain about its long-term impact on software quality. Future work should go beyond static complexity metrics and examine how AI-authored code performs over time, through indicators such as defect density, review latency, test coverage, and bug-fix effort. Moreover, the effects of AI vary across project types: infrastructure libraries, mobile apps, and backend systems may each exhibit different tolerances for simplicity vs. optimization. This calls for more nuanced evaluations that factor in domain, project scale, and team practices.

To support this direction, we release our curated \dataset{} dataset of \agentprs, alongside full histories of each developer's contributions, including pre ``\agenticais'' periods. This enables benchmarking across time, authorship, and tool usage, providing a robust foundation for studying agentic code quality. By anchoring future research in empirical evidence and open data, we aim to move beyond anecdotal impressions of AI coding agents and toward a systematic understanding of when, where, and how these agents produce code that truly meets, or exceeds human craftsmanship and standards. This is especially critical as generative models are deployed in real-world, high-stakes environments where quality cannot be compromised.

%

\researchdir{Analyze and improve the planning phase of human-AI software collaboration.} As AI agents become more autonomous in software development, one key but often overlooked phase is how tasks are scoped and planned before any code is written. This planning process is central to translating human intent into executable steps, and it plays a defining role in the success of AI-generated code. However, many current agents obscure this reasoning stage, making it difficult to understand how decisions are made or how errors originate. Tools like \copilot preserve rich planning traces embedded in PR descriptions. For instance, in a bug-fix PR,\footnote{\url{https://github.com/owncast/owncast/pull/4354}, see the edit history of PR body} \copilot automatically generated a multi-step action plan that was updated iteratively across five versions, demonstrating how it contextualizes an issue and proposes structured solutions. These traces offer a valuable research opportunity: future work could investigate what kinds of input prompts and planning structures lead to more effective outcomes, how developers interact with AI-generated plans, and what planning capabilities differentiate stronger agents from weaker ones. As task planning becomes a defining component of agentic workflows, studying it in depth will be crucial for improving reliability, transparency, and overall software quality.

\researchdir{Investigate how programming language features influence agent effectiveness and reliability.}
For example, an open question is whether languages with stronger type systems—such as TypeScript and Rust—offer implicit advantages to autonomous agents. These languages may provide a form of built-in scaffolding via richer compile-time feedback and structural constraints, helping agents both avoid errors and converge on correct implementations more efficiently. This hypothesis points to a promising area of future research at the intersection of programming language design and AI-powered automation: how language-level affordances can support, amplify, or constrain agentic behavior. Deeper study of this interplay could guide both the selection of languages that better support agent capabilities and shape the design of language-aware AI teammates.




\section{The Road Ahead}\label{sec:discuss}



\dataset{} demonstrates that autonomous agents have already moved beyond proof-of-concept prototypes and are now embedded in everyday software development. In this section, we look ahead to how \agenticais may reshape the future of software engineering, and we offer a roadmap for research and practice.

\subsection{Software Repositories as Training Environments for AI Agents}\label{sec:repo-gym}

\agenticais differ fundamentally from traditional coding assistants~\cite{hassan_aiware_2024}. Rather than simply completing code snippets in response to a developer's keystrokes~(e.g., auto-completion), they can execute entire development tasks autonomously~\cite{applis2025unifiedsoftwareengineeringagent}. For example, \agenticais can set up environments, draft implementation plans, execute tests, and submit PRs. This shift from reactive assistance to proactive agency introduces a new paradigm: treating software repositories as interactive agent training environments.

We propose that the research community embrace software repositories as structured \textit{environments}, analogous to reinforcement learning (RL) simulators such as OpenAI Gym~\cite{brockman2016openaigym}, in which \agenticais can iteratively improve through interaction. Each software repository offers natural \textit{reward} signals: successful CI builds, merged pull requests, or positive reviewer feedback. In contrast, failed tests or rejected PRs yield negative feedback. Unlike synthetic benchmarks with hand-crafted and often fragile reward functions, repositories provide feedback grounded in real-world human judgment and developer expectations. Two key observations motivate this: (i) Pretraining LLMs on code has been shown to improve general reasoning capabilities, with recent work reporting gains of up to 8.2\% over text-only pretraining~\cite{aryabumi2025to}, reinforcing the dual role of software repositories as both SE benchmarks and enablers of broader AI capabilities. (ii) As AI Agents inherently leverage the test time scaling~\cite{snell2025scaling} of LLMs, software repositories naturally serve as ``learning environments,'' enabling agents to continually improve through experience, in line with RL-style paradigms~\cite{silver2025welcome}.

\subsection{Dynamic Benchmarking and Living Leaderboards}\label{sec:benchmarks}

Traditional benchmarks such as HumanEval~\cite{chen2021evaluating} and SWE-bench~\cite{jimenez2024swebench} offer useful insights into AI performance, but they fall short in capturing the evolving, collaborative, and context-rich reality of real-world software projects. These benchmarks are static, limited to single-language (typically Python) tasks, and often lack important aspects such as code review dynamics or evolving team preferences. Moreover, they are increasingly susceptible to data contamination~\cite{liang2025swebenchillusionstateoftheartllms} and leakage through pretraining~\cite{zhou2025lessleakbenchinvestigationdataleakage}, which undermines their ability to reflect generalization and real-world applicability.

We advocate for a shift toward dynamic benchmarking using live data from active repositories. Agentic contributions can be assessed on an ongoing basis using metrics such as acceptance rates, review turnaround times, and code quality indicators. This approach enables the development of a living leaderboard, similar to the Chatbot Arena~\cite{chiang2024chatbotarenaopenplatform}, but tailored for software engineering agents. Such a leaderboard would reflect not only who solves the most test cases, but also who integrates most effectively into real projects.

Recent work has begun to move in this direction. For example, SPICE~\cite{bhatia2025spice} proposes a modular benchmarking pipeline and introduces a manually curated SWE-bench-style dataset for evaluating AI-generated patches across a wide range of open source projects. Notably, SPICE is fully open-sourced, providing both tools and datasets that the community can adapt to their own evaluation needs. It also offers valuable commentary on the high cost—both manual and computational—of building reliable, real-world benchmarks. These insights reinforce that while such efforts are essential, they are also resource-intensive and require careful design tradeoffs.

Ultimately, these efforts highlight that effective benchmarking in the age of AI teammates must go beyond static, general-purpose leaderboards. We anticipate a growing need for customized benchmarks that reflect the unique characteristics, workflows, and quality standards of individual projects or organizations. Dynamic and domain-specific benchmarking supports more meaningful comparisons between agents, enables informed decision-making for tool integration, and better aligned evaluation metrics with the realities of real-world software engineering practice.

\subsection{Towards SE 3.0 Methodologies}\label{sec:se30}

Agile, Scrum, and DevOps practices emerged in a world where all contributors were human, working in diurnal cycles and constrained by cognitive and communication limits. \agenticais fundamentally disrupt these assumptions~\cite{Qiu2024FromTCC,Lin2024SOEN101CGD}. A single developer can now orchestrate multiple autonomous agents, each capable of independently creating pull requests, writing tests, generating documentation, and responding to review feedback around the clock~\cite{Dong2023SelfCollaborationCG, Bouzenia2024RepairAgentAA}. This shift alters the cadence of development, the locus of decision-making, and the definition of team velocity~\cite{He2024LLMBasedMSF, roychoudhury2025agenticaisoftwareengineers}. The influx of agentic contributions also introduces new challenges. How should developers manage the review load from always-active agents? How should teams triage overlapping changes or resolve the conflicted versions? These complexities call for new software engineering methodologies.

We propose that SE 3.0 requires a rethinking of practices and infrastructure: (1) Orchestration frameworks are needed to manage high-throughput agentic workflows while preserving human oversight and control. (2) Review systems must adapt to triage contributions from \agenticais, potentially requiring agents to provide structured rationales, test plans, or explanations for code edits. (3) Governance models should clarify when agents are allowed to act autonomously, how their contributions are validated, and who holds accountability for defects. Just as Agile reshaped engineering for development, SE 3.0 must redefine engineering practices for a world in which autonomy and collaboration are not mutually exclusive, but deeply interwoven. Hybrid human-AI development teams will demand new norms, tools, and theories. This is an opportunity for the research community to lead this methodological transition.

\section{Conclusion}\label{sec:conclusion}

We present \dataset{}, a large-scale dataset capturing the real-world activities of Autonomous Coding Agents within GitHub pull requests. While prior work has speculated about the emergence of AI-native software development (SE 3.0), \dataset{} offers the first empirical and systematic evidence of how widely used \agenticais such as OpenAI Codex, GitHub Copilot, Devin, Cursor, and Claude Code contribute to modern codebases. The dataset contains over 456K PRs authored by AI Agents across 61K GitHub repositories and 47K developers.

Our analyses reveal surprising and sometimes sobering insights into the integration of \agenticais into real-world workflows. For example, although these agents can dramatically accelerate code contribution rates, substantial disparities remain in PR acceptance rates and review dynamics when compared to human developers. Notably, agents like \codex can close PRs up to ten times faster than humans, yet quality and trustworthiness remain open challenges. \agentprs are significantly less likely to be merged, particularly when they involve complex reasoning tasks such as feature development or bug fixing. These results expose a growing gap between benchmark performance and real-world effectiveness, calling into question the ecological validity of current static evaluation methods.

\dataset{} not only grounds the next wave of software engineering research in reality, but also unlocks new directions for studying productivity, authorship, code review practices, and human-agent governance. By releasing AIDev as a living, extensible resource, we invite the community to iterate on and extend this foundation gy building new benchmarks, crafting better collaboration paradigms, and advancing AI-native tooling. The future of software engineering will be shaped not just by what AI teammates can do, but by how we choose to evaluate, guide, and partner with them. \dataset{}  takes us from speculation to evidence—from theory to practice—in defining that future.

\bibliographystyle{ACM-Reference-Format}
\bibliography{main}

\end{document}